\documentclass[10pt]{iopart}
\usepackage{latexsym}
\usepackage{epsfig}
\usepackage{epstopdf}

\usepackage{iopams}  
\begin{document}

\title{Analysis of the Schmidt, Cohen  \& Margon (1980)  features in the Red Rectangle nebula}


\author{Fr\'ed\'eric Zagury }

\address{Institut Louis de Broglie, 23 rue Marsoulan, 75012 Paris, France}
\ead{fzagury@wanadoo.fr}
\begin{abstract}
This study investigates the relationship between atmospheric extinction and the spectrum of the Red Rectangle nebula  on scales of a few to a few tens of \AA.
It is found that the fine structure of the nebula's  continuum short-ward of 6700~\AA\ is similar to background spectra, and is thus determined either by atmospheric absorption or by light from HD44179 scattered in the earth atmosphere. 

\end{abstract}
\vspace{2pc}
\maketitle

\section{Introduction}
\begin{figure*}
\resizebox{1.3\columnwidth }{!}{\includegraphics{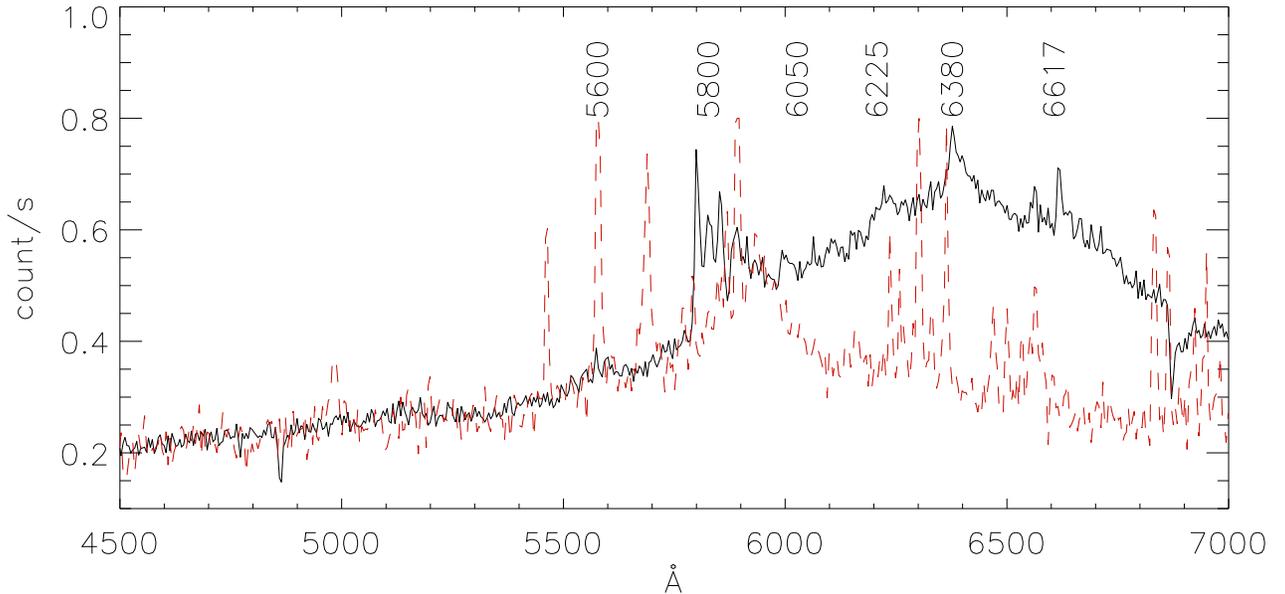}} 
\caption{Average spectrum ($s46$ to $s50$), background subtracted, of  the Red Rectangle nebula ($14''$ north from HD44179) and  the background spectrum (red dashes) taken at the edge of the slit, re-scaled by a factor of 6, in the same observation.
The six SCM80 bands are indicated.
Adapted  from fig.~11 in  \cite{rr}.
} 
\label{fig:fig1}
\end{figure*}

The first detailed spectrum of the Red Rectangle nebula was published in \cite{schmidt80} (SCM80, the original spectrum is similar to the plain line spectrum of Fig.~\ref{fig:fig1}).
It shows, over the  continuum in the blue region of the spectrum,  a broad-band bump (long-ward of $5500\,\rm\AA$, Fig.~\ref{fig:fig1})  which later on received the name of 'ERE' (Extended Red Emission) bump.
SCM80 finds structure in the bump, mainly concentrated  in six wavelength regions (see  Fig.~\ref{fig:fig1}).
At a few \AA\ resolution three of these Red Rectangle bands are diffuse, unresolved features centered at $5600$, $ 6050$ and $ 6225$~\AA, while the three others ($\lambda 5800$, $\lambda 6380$, $\lambda 6617$) are composed of one or a few  sharp lines.
Since their discovery, the origin of the Red Rectangle bands, attributed to emission by some not yet identified molecules\footnote{\emph{'We conclude that, on the basis of available information, the peculiar spectrum of the Red Rectangle cannot be reproduced by emission from a known molecular species.'} SCM80, p.~L136}, is a mystery.

In a preceding study \cite{rr} I have shown the correspondence (see Fig.~\ref{fig:fig1}) which exists between the continuum in the blue region of the Red Rectangle spectrum and background spectra.
The continuum underlying the bump, directly observed short-ward of $\sim5700\,\rm\AA$, and previously attributed to light from HD44179 (the star illuminating the nebula) scattered in the nebula \cite{schmidt80, witt90}, is, in fact, either diffracted light from HD44179, for small angular distances from the star, or, at larger distances, a mixture of light from HD44179 scattered in the earth atmosphere and night sky.
In the blue, light from the nebula is thus a negligible part of the spectrum, much fainter than the background.
Only the red rise of the bump can safely be attributed to the nebula and  indicates Rayleigh scattering by gas (presumably hydrogen), in conformity with the proximity of the illuminating star and the large angles of scattering.
The blue side of the ERE bump (5500 to 6500~\AA),  in-between the blue atmospheric background and the red rise, appears as a transition region.
It coincides with, and may be affected by the strong  ozone  Chappuis absorption bands.

 \cite{rr} was restricted to the study of the broad band shape of the nebula's spectrum but these conclusions  question the nature of its fine structure, and, maybe, some of the Red Rectangle bands.
In this respect, several figures of  \cite{rr}  show resemblance between nebular and background spectra  on scales of a few \AA.
This appears more evidently on Fig.~\ref{fig:fig1} where a background spectrum  (red dashed spectrum of the figure) has been re-scaled to match the spectrum of the nebula in the blue.
In the wavelength region of the first Red Rectangle band ($\lambda 5600$) for instance,  the spectra superimpose perfectly well,  suggesting that this band has an atmospheric origin.

My purpose here is to look again at the data that I used in  \cite{rr} (and for Fig.~\ref{fig:fig1}), which have a  slightly better resolution than SCM80's, with the aim of analyzing the relationships which exist between the spectrum of the nebula and the background on scales of a few \AA, and to see the implications of such relations for the comprehension of the SCM80 Red Rectangle bands.
This comparison will be supported by higher resolution data retrieved from the ESO Archive Data Center facility.
 
The first part of the article is a factual analysis of the data.
I first recall and complete a few results obtained in  \cite{rr} and in \cite{fd}, on the structure of the background spectrum in the vicinity of a star (Sect.~\ref{fd}).
Sect.~\ref{neblim} analyzes the variations of the spectra along the slit in the Red Rectangle observations.
The comparison of background and nebular spectra over small wavelength intervals covering the SCM80 features is done in Sect.~\ref{cfn}.
 
The second part of the article (Sect.~\ref{dis}) discusses implications  of this comparison for the Red Rectangle spectrum, the pattern of the night sky, especially in the 5800~\AA\ region, and the correction of atmospheric extinction by standard data reduction routines.
\section{Data} \label{data}
\subsection{Red Rectangle data} \label{data-rr}
The study relies on a large series of spectra of the Red Rectangle nebula obtained  between 2001 and 2003 with the FAST long-slit spectrograph \cite{fabricant} mounted on the 1.5~m Tillinghast telescope of the Fred L. Whipple Observatory at Mount Hopkins (USA).
These data were analyzed  in  \cite{rr}.

Observations of the nebula were made at different distances, $5''$ to $14''$ from HD44179, and repeated over the years. 
The resulting 2-D array of an observation covers the [3660, 7530]\AA\  range with 1.4\,\AA\ spectral resolution and $1.2''$ spatial resolution.
Slit dimensions are $3'' \times 3'$. 

FAST/Mount Hopkins data are complemented by higher resolution spectra  retrieved from  the  ESO/ST-ECF Science Archive Facility  (http://archive.eso.org/) and observed  with the 3.5~m NTT telescope at La Silla-Paranal Observatory (Chili).
Ten long-slit spectra in the Red Rectangle were obtained in January 25 and 26, 1998, with spectro-imager EMMI (ESO Multi Mode Instrument) and  two gratings: grating 6  (spectral resolution: $0.32\,\rm\AA$; spectral coverage: $[5540, 6195]\,\rm\AA$) and grating 7 (spectral resolution: $0.65\,\rm\AA$; spectral coverage: $[5520, 6880]\,\rm\AA$). 
The slit is $1''$ wide, $3'$ long on the sky.
Outputs of these observations are 2-D arrays with 700 pixels along the spatial dimension ($0.268''$ per pixel spatial resolution) and 2086 pixels along the spectral one.
These ESO observations are describe in \cite{winckel02}.

Red Rectangle spectra in the figures of this article  come from the following observations:
\begin{itemize}
\item Mount Hopkins FAST observation of December 2001 (exposure time 240~s), 14'' north from HD44179.
\item ESO data-set ONTT.1998-01-26T01/07/07.550.fits, observed  $6''$ north from HD44179 with grating~6, an observing time of 2700~s, and P.A.= $105^{\circ}$.
This observation corresponds to cut C on figure~1 of \cite{winckel02}.
\item  ESO data-set ONTT.1998-01-26T01/58/37.630.fits: $11''$ north from HD44179, grating~7, 3600~s exposure time, P.A.= $105^{\circ}$, cut D in \cite{winckel02}.
\item ESO data-set ONTT.1998-01-25T03/30/22.510.fits:  grating~6, 2400~s exposure time,  P.A.= $45^{\circ}$, the slit follows the north-eastern rim of the nebula (cut A in \cite{winckel02}).
\end{itemize}
The spectrum at pixel `$x$' in a 2-D array of these long-slit observations will be noted `$sx$'.
The 'main' pixel or spectrum of a 2-D array designates the pixel/spectrum with maximum signal on the slit.
Figures will, in general, present raw data, flat-fielded but not corrected for atmospheric extinction.
Except for Fig.~\ref{fig:fig1}, no background is subtracted from the spectra of the nebula.
Units (count/s) depend on the spectrograph (FAST or EMMI) and its configuration (for EMMI observations). 

The spectrum of a star, as HD44179, in either NTT/EMMI or Tillinghast/FAST observations, extends over several pixels along the spatial direction.
For a given wavelength it is well represented by a gaussian  which for NTT/EMMI observations has  $\sim 8$~pixels ($2''$) full width at half maximum, and  $\sim 3.2$~pixels ($3.4''$) for the Tillinghast/FAST ones.
I  usually refer to this direct light from HD44179 as 'direct' or 'diffracted' starlight.
The wings of the gaussian are observed relatively far-away from the star: in the observations of the Red Rectangle nebula it gives the most important contribution to the background 5'' from HD44179 (fig.~9 in  \cite{rr}) and fades out under starlight  scattered in the atmosphere between $5''$ and $10''$.

\subsection{Additional astronomical data} \label{data-ast}

In addition to the Red Rectangle data, I will also use a night sky spectrum at Kitt-Peak observatory \cite{massey00}  provided by P.~Massey.
\subsection{Atmospheric data}  \label{data-atm}
Atmospheric data used for Sect.~\ref{dis} consist in occultation spectra of the sun and absorption spectra of a few atmospheric molecules.

Occultation spectra are based on observations by  balloon borne experiment SAOZ (Syst\`eme d'Analyse par Observation Z\'enitale, \cite{saoz}), and were analyzed in  a previous paper \cite{sol1}.
The spectra correspond to observations of the sun through atmospheric layers of increasing thickness.
They are normalized by the unreddened spectrum of the sun (obtained at the beginning of the observations, when the sun was above the atmosphere). 
Resolution is $6.33\,\rm\AA$.

Atmospheric absorption is a complex domain of research: theoretical calculations are generally difficult to handle owing to the complexity of atmospheric molecules, absorption cross-sections from laboratory experiments can be found in a few cases but often lack resolution or do not cover the whole visible spectrum.
In addition to column density, absorption spectra are sensitive to pressure and temperature.
There is, since several years, an effort to gather experimental data into a coherent data-base (HITRAN for HIgh-resolution TRANsmission molecular absorption; \cite{rothman05, rothman98}; http://cfa-www.harvard.edu/hitran/).
HITRAN gives absorption line positions and main parameters which can be converted into absorption spectra.

Atmospheric absorption spectra in this paper are derived  from the HITRAN database or from absorption cross-sections  found on the websites of the Molecular Spectroscopy and Chemical Kinetics group at the University of Bremen (Germany) and the British Atmospheric Data Centre (http://badc.nerc.ac.uk/data/msf/).
A high resolution  (0.01~\AA) laboratory spectrum of NO$_2$, due to A. Jenouvrier (Universit\'e de Reims, France), will also be used. 
Absorption spectra  for ozone, OH,  O$_4$, and H$_2$O presented here have respective resolutions of
0.2~\AA, 0.25~\AA, 0.56~\AA, and 0.23~\AA.
\begin{figure}[h]
\resizebox{\columnwidth }{!}{\includegraphics{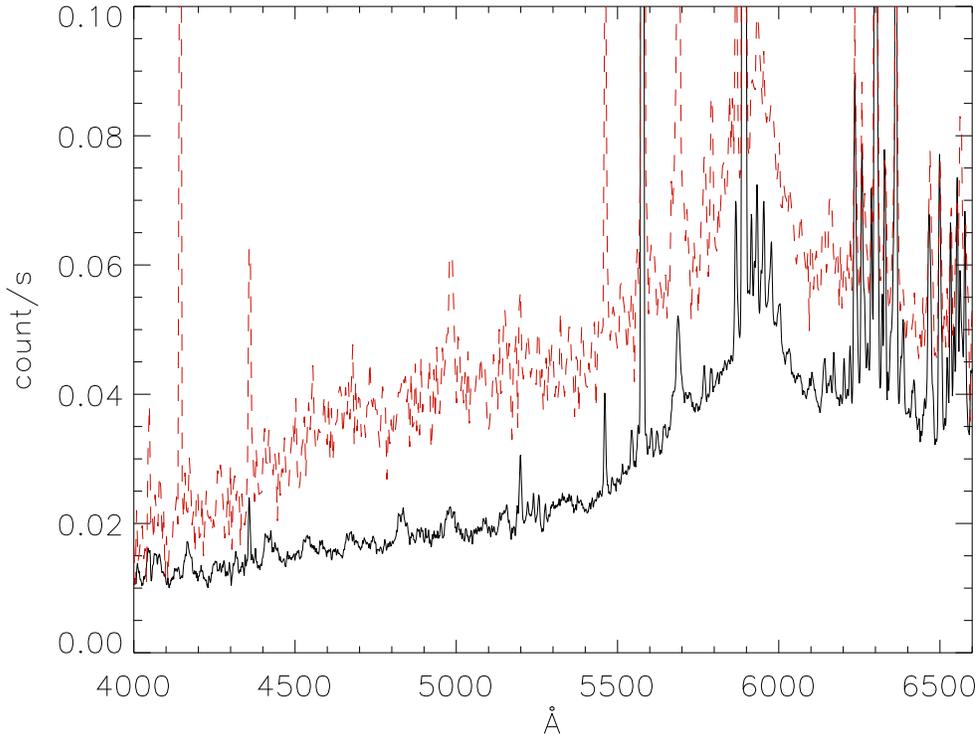}} 
\caption{Background spectrum (red) in the FAST observation of the Red Rectangle nebula and the night sky spectrum of \cite{massey00}.
The background is the observed spectrum, outside the nebula,  $\sim 55''$ from HD44179.
} 
\label{fig:fig2}
\end{figure}
\section{Night sky and backgrounds} \label{fd}
\subsection{Kitt Peak and Mount Hopkins night-skies} \label{fdgene}
Night sky spectra in the visible, at  Kitt Peak and Mount Hopkins observatories, are describe in  \cite{massey00,massey90}.
They have an identical shape (figs.~1, 2 in \cite{massey00} and Fig.~\ref{fig:fig2}), independent of direction (figs.~1a and 1b in \cite{massey90}) and constant over the years \cite{massey00,massey90}, except maybe in the $5800\,\rm\AA$ region (fig.~2 in \cite{massey00}).

From the blue to the red, the shape of the night sky spectrum consists  in a rise (a plot as function of wave-number shows it is the tail of an exponential), a bump over-which the atmospheric NaD lines (eventually also city light, see Sect.~\ref{ns}) are superimposed (between $5800\,\rm\AA$ and $6100\,\rm\AA$), and two sets of  emission lines,  centered at $ 6300\,\rm\AA$ (OH and OI mainly) and $6550\,\rm\AA$ (OH).

\subsection{Modification of the  night sky spectrum in the vicinity of a star} \label{fdet}
Diffraction and scattering in the atmosphere can modify the night sky spectrum several arc-minute around the direction of a star, depending on the star magnitude \cite{fd}.

In the immediate vicinity (first few arc-second) of the star the observed spectrum is dominated by its direct (diffracted) light  (Sect.~\ref{data-rr}), which is, as angular distance $\rm\theta$ from the star increases, rapidly replaced by  scattered starlight in the atmosphere, and  night sky.

Scattered starlight spectrally behaves as $1/\lambda$ times the spectrum of the star (figs.~3, 7 and 11 in  \cite{rr}; fig.~11 in \cite{fd}), and is due to aerosols.
Its decrease with $\rm\theta$ is steep in the immediate vicinity of the star, slow at larger distances (a $1/\rm\theta^2$ dependence should be expected).

The spectrum observed in the vicinity of a star is therefore complex and rapidly varying, both in shape and magnitude.
Its magnitude with increasing $\theta$ quickly becomes  negligible compared to the magnitude of the star but remains, over a few ten arc-seconds, larger than the night sky spectrum and  modifies its shape (Fig.~\ref{fig:fig2}). 

If the star illuminates a nebula,  direct light from the star can considerably enhance and alter  the observed spectrum of the nebula over a few arc-seconds around the star direction.
At larger distances, scattered starlight in the atmosphere and night sky underly the spectrum of the nebula.

The background to be subtracted to the observed spectrum in the observation of a nebula is thus generally not the spectrum observed far away from the nebula (at the edges of the slit in long-slit spectroscopic observations), may significantly differ from it, and  be difficult to determine.
\subsection{Background in the FAST/ Mount Hopkins observations of the Red Rectangle nebula} \label{fdneb}
\begin{figure*}
\resizebox{1.3\textwidth }{!}{\includegraphics{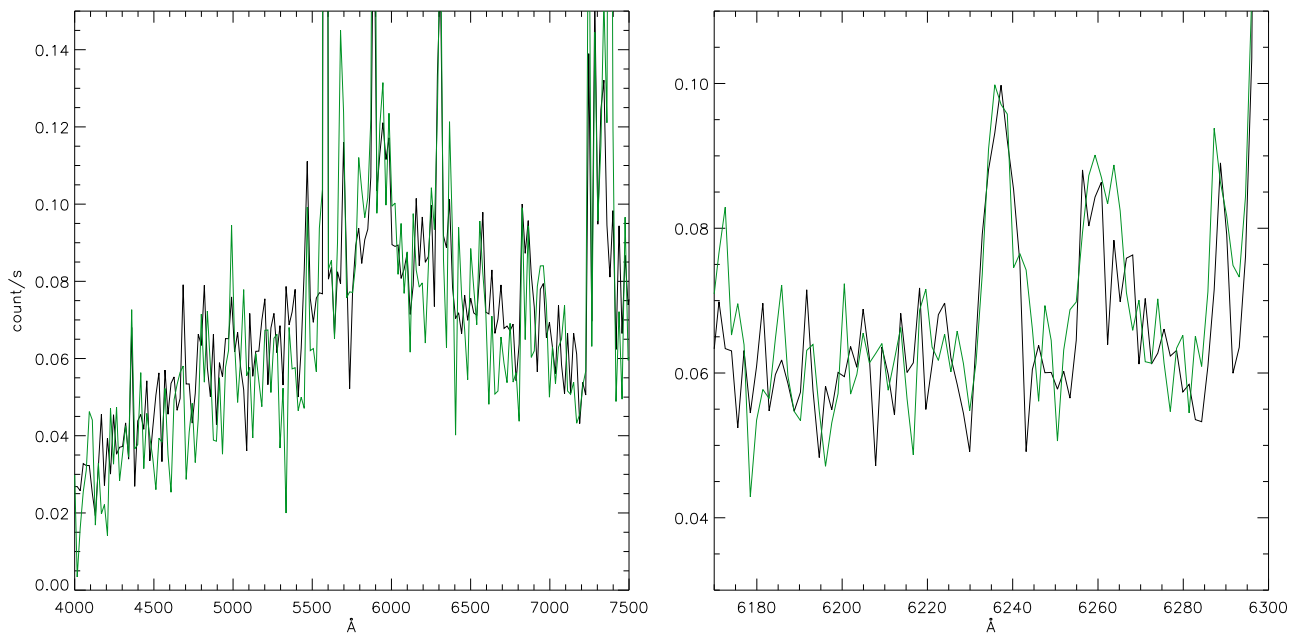}} 
\caption{The background in the FAST observation of the nebula  ($14''$ north from HD44179).
\emph{Left plot:} $s35$ (black spectrum) and $1.4s0$ (in green), both with a 13 points median filter, are almost identical: the average level of the background decreases towards the edges of the slit but the structure remains the same.
\emph{Right plot:} comparison of the fine structure of two background spectra (averages of $s10$ to $s30$, in black, and of $s70$ to $s90$, in red) at each edge of the slit, in the $[6180,\,6300]$~\AA\ wavelength range.
} 
\label{fig:fig3}
\end{figure*}
\begin{figure*}
\resizebox{1.3\textwidth }{!}{\includegraphics{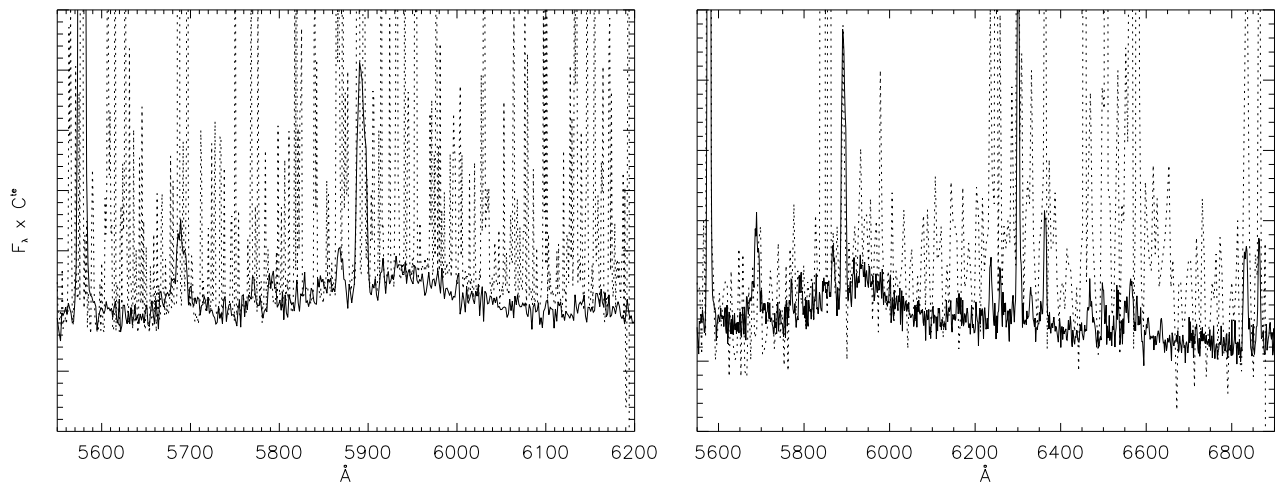}} 
\caption{The figure shows that Mount Hopkins and La Silla-Paranal backgrounds in the observations of the Red Rectangle nebula are similar. 
Re-scaled Mount Hopkins background (plain line) is compared to La Silla backgrounds in the observations $11''$ from HD44179 (right plot), and to the observation with P.A.=45$^{\circ}$ (left panel).
} 
\label{fig:fig4}
\end{figure*}
$5''$ from HD44179 diffracted light underlies the spectrum of the nebula  (fig.~9 in  \cite{rr}).
It is typically a factor of ten larger than starlight scattered in the atmosphere and has a much different shape   (fig.~2 in  \cite{rr}).
It explains the high level of the Red Rectangle nebula spectra observed at very short  distances from HD44179 and is the probable reason for the 'dramatic' change in the spectrum mentioned in \cite{winckel02,glinski02}, between observations $5''$ and $10''$ from HD44179:
it is indeed between $5$ and $10''$ from HD44179 that scattered starlight in the atmosphere progressively replaces diffracted light.

The relative importance of diffracted and scattered lights from HD44179 in the spectrum of the nebula, for small $\theta$, depends not only on $\rm\theta$ but also on air-mass and atmospherical conditions: the same observation ($5''$ from HD44179, fig.~24 in  \cite{rr}) repeated at different dates can show important modifications in the continuum of the spectrum of the nebula.

$10''$ from HD44179 scattered starlight in the atmosphere is roughly a factor of ten above the night sky spectrum.

At distances larger than $\sim 10''$ from HD44179, the background spectrum consists of scattered starlight upon which the night sky spectrum is superposed (Fig.~\ref{fig:fig2};  figs.~7, 8 in  \cite{rr}).
Left plot of Fig.~\ref{fig:fig3} shows that, at such distances, the background decreases with distance from HD44179 with little modification of its shape.
Right panel of the figure proves that the fine structure of the background in the FAST observations remains meaningful down to a few $\rm\AA$.
\subsection{La Silla observations} \label{fdntt}
As for Mount Hopkins observations, the background in ESO observations depends on and varies with  distance from HD44179.

It is remarkable to see that the background far-away from HD44179 reveals a similar shape as at Mount Hopkins (Fig.~\ref{fig:fig4}).
It implies that the characteristic pattern of the night sky spectrum found in \cite{massey90} is, in the visible, identical at Kitt-Peak, Mount Hopkins, and La Silla.

As for the FAST observations,  variations of the background spectra in these observations remain  meaningful down to scales of a few \AA.
\begin{figure*}
\resizebox{1.3\textwidth }{!}{\includegraphics{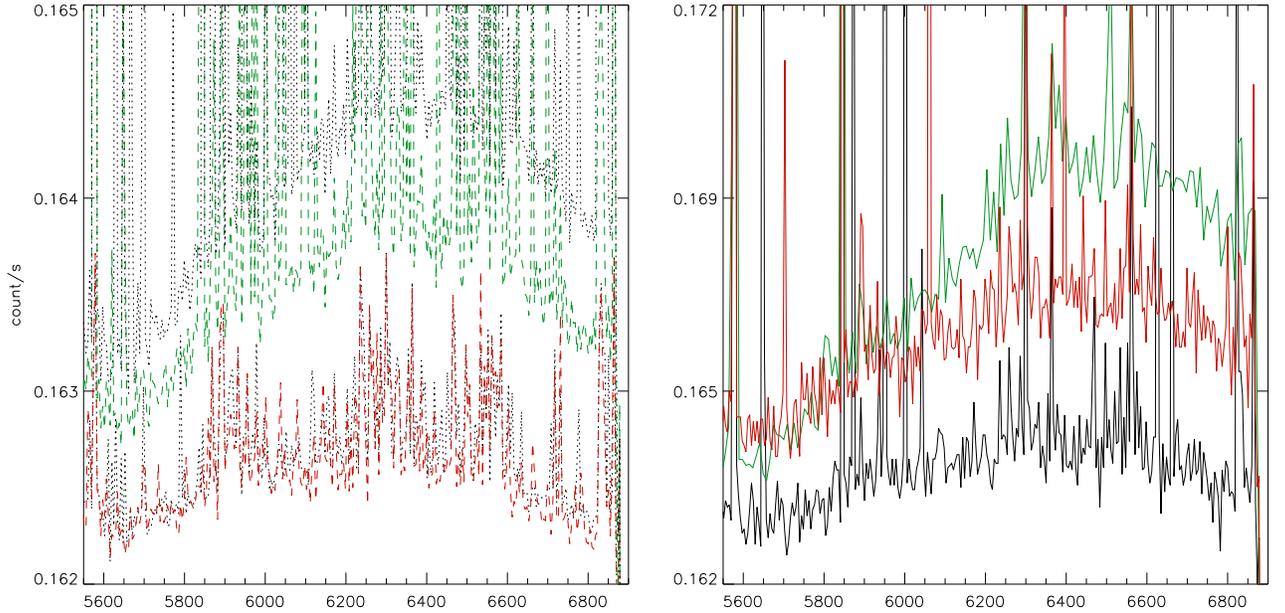}} 
\caption{Backgrounds in ESO observation of the Red Rectangle nebula, $11''$ from HD44179.
\emph{Left:} variations of the background along the slit: the two bottom spectra are the average of $s670$ to $s690$ (red dashes) and $s540$ to $s560$ (black dots). 
Medium and top spectra are the average of $s450$ to $s460$ (green dashes) and of $s440$ to $s450$.
All these spectra are nearly proportional: they are a mixture of night sky and HD44179's light scattered in the atmosphere.
\emph{Right:} from bottom to top, $s450$ (in black, with a thirteen  points median filtering), $s435$ (red, shifted by $+0.01\,count/s$,  thirteen  points median filtering), $s427$ (green, with a seven points median filtering).
The spectra have a common structure, but there is an increase of brightness in the red for $s427$, which may be interpreted as the appearance of the nebula's spectrum.} 
\label{fig:fig5}
\end{figure*}
\section{Limits of the nebula} \label{neblim}
Fig.~\ref{fig:fig5} plots several background spectra along the slit of ESO's observation $11''$ from HD44179.
Backgrounds are nearly undistinguishable (Fig.~\ref{fig:fig5}, left plot, bottom spectra) up to $\sim s210$ and $\sim s510$ on each side of the slit.
From there on the spectra begin to rise, but keep a similar shape (they remain close to proportionality), at an accelerating rate towards the center of the slit (Fig.~\ref{fig:fig5}, left plot). 

These first modifications of the spectra along the slit are due to the growing importance of scattered (by the atmosphere) starlight in the background: a nebular extent from $s210$ to $s510$ would make the nebula  $\sim 300$ pixels, $70''$ wide, which is unrealistic for a distance of eleven arc-second  from HD44179.

The only available way to identify the nebula on the slit is a change in the red, which occurs  (Fig.~\ref{fig:fig5}, right plot) at $\sim s330$ and $\sim s429$ and yields a width of $\sim 27''$.
Exact width might be slightly smaller, by a few arc-seconds, because of the spread of light from one direction over several pixels of the slit (Sect.~\ref{data-rr}).
The same criterion applied to the FAST observation $14''$ from HD44179 gives a $\sim 31''$ (from pixels 38 to 63) width.

At close angular distance from HD44179 the problem of fixing the limits of the nebula on the slit is worse since  diffracted light from HD44179 is predominant and affects the red slope of the spectrum.

\begin{figure*}
\resizebox{1.3\textwidth }{!}{\includegraphics{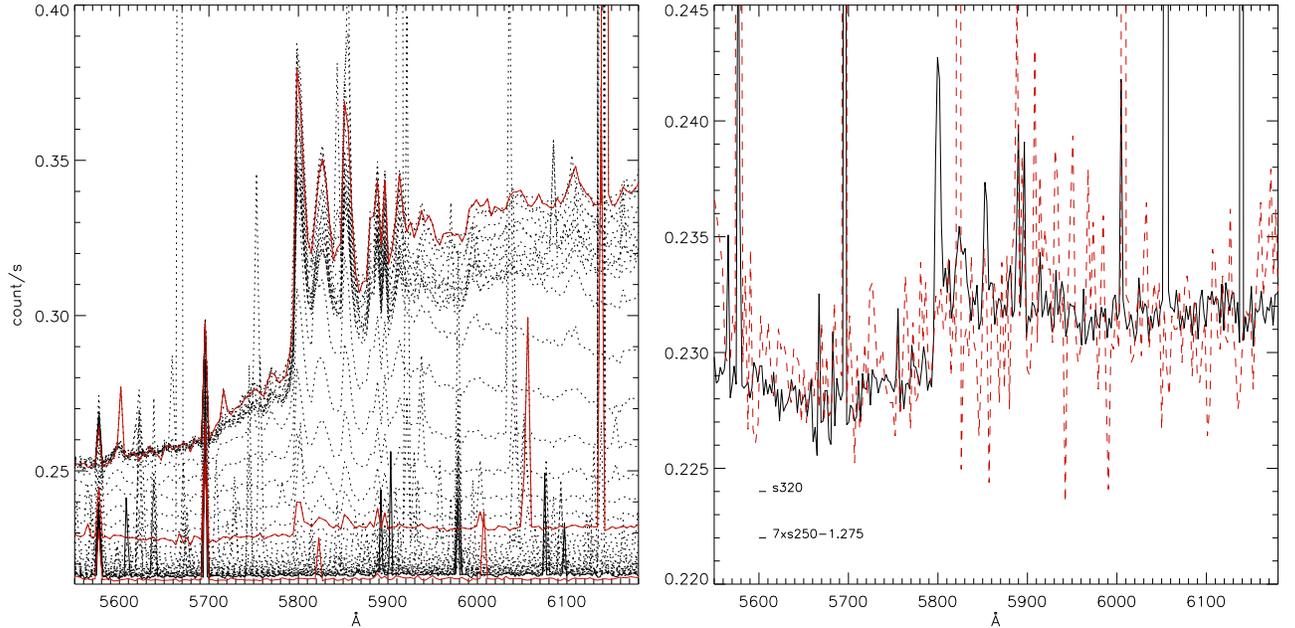}} 
\caption{ESO observation of the Red Rectangle nebula $6''$ from HD44179.
\emph{left:} Spectra $s250$ (very bottom  spectrum), out of the nebula, spectra $s331$ (main spectrum of the 2-D array, top curve) and $s320$,  in the nebula, are in red.
Dotted spectra are all other spectra ranging from $s295$ to $s340$.
All spectra have been convolved by a 13 point median filter.
The plot shows the continuity of change of shape between the background and $s331$, thus the difficulty of fixing the limits of the nebula.
\emph{right:} Strong similarities remain between the shapes of spectra $s320$ (solid line), within the nebula, and  $s250$ (times 7 and shifted by -1.275, in red), outside the nebula.
} 
\label{fig:fig6}
\end{figure*}

In ESO observation $6''$ north from HD44179 (main spectrum $s330$) backgrounds are all proportional from $s0$ to $\sim s295$ and from $\sim s400$ to $s699$ (Fig.~\ref{fig:fig6}, left plot).
A change of shape on individual spectra appears in the $5800\,\rm\AA$ region at pixels 295 and 400 on each side of the slit (Fig.~\ref{fig:fig6}, left plot).
This would mean a $\sim 106$ pixels, $\sim 28''$ extent,  as found $11''$ from HD44179, and larger than was reported in earlier papers  ($\sim 9''$ in \cite{winckel02}, from the same observation).
In addition one notes a persisting relationship between the overall shape of spectra  in and outside the nebula  (Fig.~\ref{fig:fig6}, right plot).
\begin{figure*} 
\resizebox{1.2\textwidth }{!}{\includegraphics{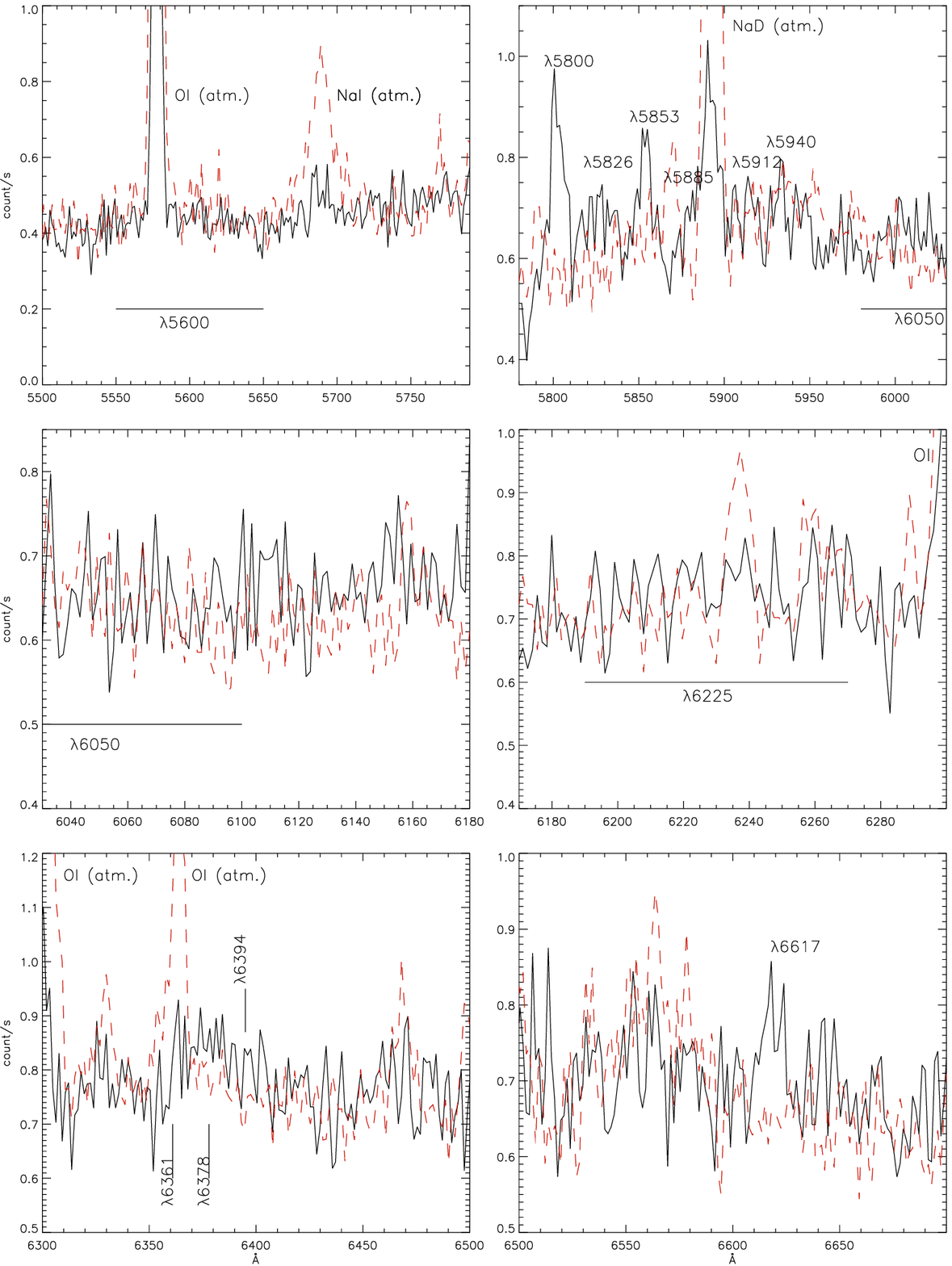}} 
\caption{Detailed comparison of the background (in red) and of $s48$ (no background subtracted), main spectrum in  the Red Rectangle observation $14''$ north from HD44179  (same observation as in Fig.~\ref{fig:fig1}) in several wavelength bands.
The background (average of $s10$ to $s30$) is multiplied by 6.7.
From left to right and top to bottom it  is offset by +0.05, 0.1, 0.2, 0.3, 0.38, 0.32.
} 
\label{fig:fig7}
\end{figure*}
\section{Detailed comparison of nebular and background spectra in the Red Rectangle observations} \label{cfn}
To investigate the relationship between nebular and background spectra at a few $\rm\AA$ scales, I have compared (Fig.~\ref{fig:fig7})  over short wavelength intervals, $s48$, the spectrum of the nebula with maximum signal on the slit in the FAST observation ($14''$ from HD44179), to the background in the same observation ($\sim55''$ from HD44179).

To highlight the similarities between the spectra, the background spectrum is multiplied by $6.7$ to scale  the amplitude of its variations with that of spectrum $s48$ in the nebula.
This re-scaling has a physical reason.
Variations of the spectra due to atmospheric absorption depend on the continuum and are thus much fainter far away from HD44179 than in the nebula.
Therefore the recognition of atmospheric absorption effects in the spectrum of the nebula requires  that the background spectrum be re-scaled in order for its variations to be comparable to the variations of the spectrum of the nebula.
Reciprocally, features which appear to be identical in the spectrum of the nebula and in the re-scaled spectrum of the background must be (unless they belong to HD44179, see Sect.~\ref{orig}) due to absorption by the atmosphere. 

In contrast atmospheric emission lines are added identically to all spectra; they appear a factor of $\sim7$ larger in the re-scaled background spectrum (see for instance the NaD emission line, right top plot of Fig.~\ref{fig:fig1}). 

An offset is then applied to the re-scaled background in each wavelength range.
This offset (0.05, 0.1, 0.2, 0.3, 0.38, 0.32~count/s, from left to right and top to bottom of   Fig.~\ref{fig:fig7})
may physically correspond to the (average) variation of the nebular spectrum between the successive  wavelength intervals.

The result, Fig.~\ref{fig:fig7}, demonstrates a surprisingly good correlation between the variations of background and nebular spectra over short wavelength intervals.
\subsection{SCM80's $\lambda5600$, $\lambda6050$, $\lambda6225$ diffuse  bands} \label{deb}
\begin{figure}
\resizebox{\columnwidth }{!}{\includegraphics{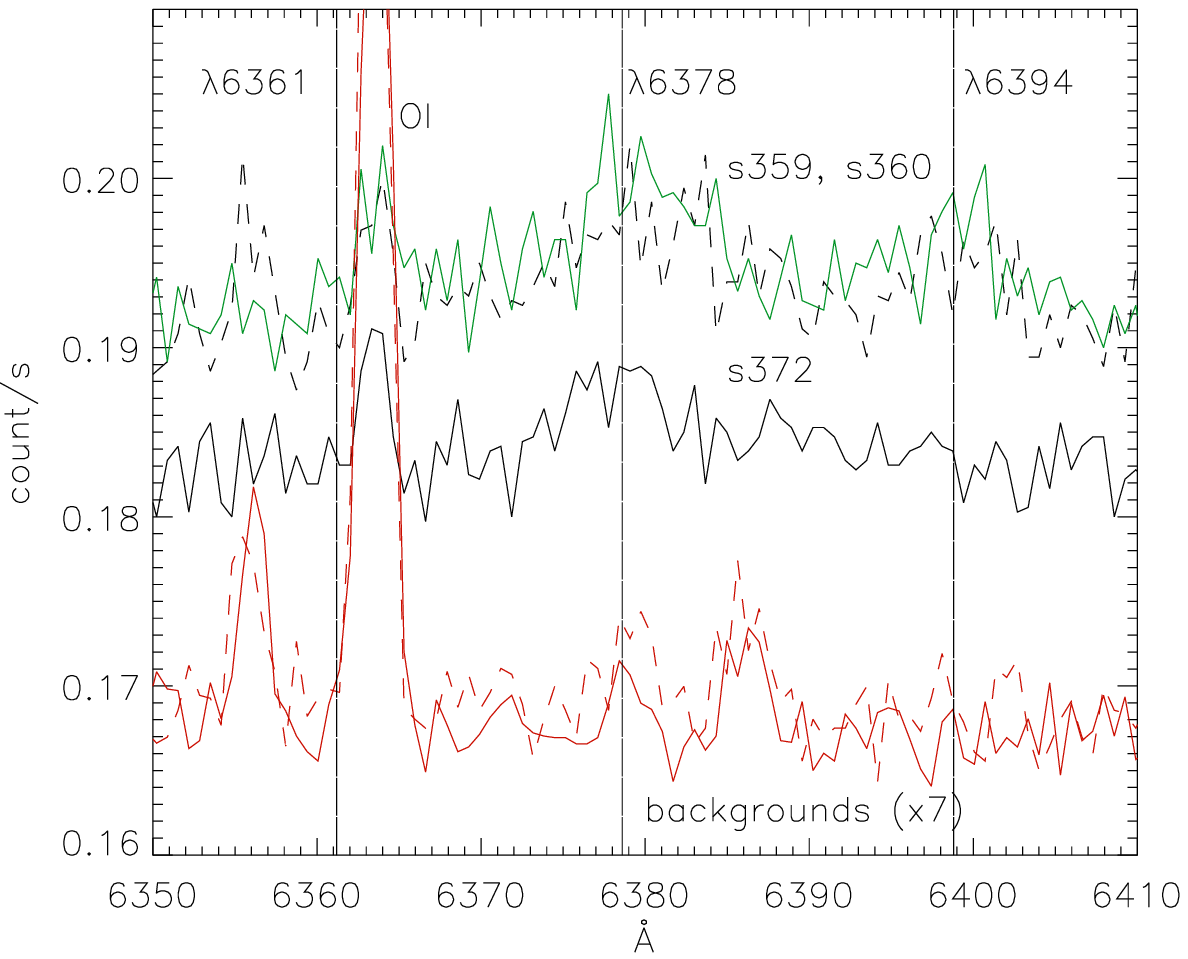}} 
\caption{The $\lambda6380$ SCM80 complex  in ESO observation of the Red Rectangle, $11''$ north from HD44179.
Central wavelengths of the three lines are indicated by vertical lines.
Three positions ($s359$, $s360$, $s372$) in the nebula, and two backgrounds (averages of  $s250$ to $s270$ and $s491$ to $s501$, in red), one on each side of the nebula, are represented.
The feature in the background between 6384 and 6388~\AA\ corresponds to two telluric emission lines listed in the La Silla sky emission atlas \cite{hanuschik03}.
Backgrounds are multiplied by 7 and  offset by -0.97, which amplifies their variations but leaves unchanged their average level.
} 
\label{fig:fig8}
\end{figure}
\begin{figure}
\resizebox{\columnwidth }{!}{\includegraphics{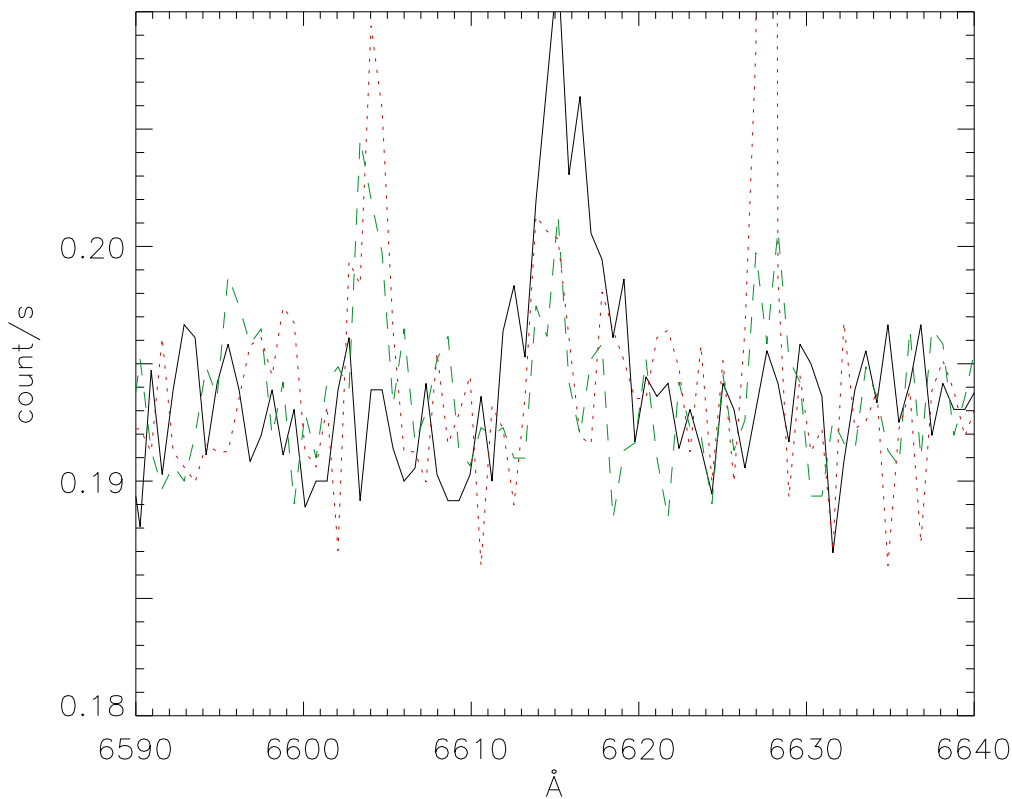}} 
\caption{$\lambda6617$ SCM80 sharp emission line in ESO observation of the Red Rectangle, $11''$ north from HD44179:
$s359$ (main nebular spectrum of the 2-D array, solid line) and averages of $s320$ to $s325$ (green dashes), $s445$ to $s450$ (red dots).
The average spectra have been multiplied by 7 and are offset by a constant.
$\lambda6617$ is present at least from $s325$ to $s445$, that is over 120 pixels on the slit,  $32''$ on the sky.
} 
\label{fig:fig9}
\end{figure}
$\lambda5600$ is $\sim150$~\AA\ large (from $\sim5520$~\AA\ to $\sim5670$~\AA, Fig.~\ref{fig:fig1}  and top left panel of Fig.~\ref{fig:fig7}) in SCM80 and in Mount Hopkins observation.
In the higher resolution ESO observations, the spectrum is nearly flat between 5550 and 5700~\AA, with only a small ($\sim50$~\AA\ large) bump between $\sim5570$ and 5620~\AA.
The analysis of this band is complicated by the proximity of two atmospheric emission lines at 5577~\AA\ (OI) and 5688~\AA\ (NaI) (Fig.~\ref{fig:fig7}).
Re-scaled background and nebular spectra are however nearly identical up to 5750~\AA.

$\lambda6050$  extends from $\sim5985$ to 6090~\AA.
It is not clear to me (see figs.~1 and 2 in \cite{winckel02} and Fig.~\ref{fig:fig1}) whether this band should be considered as a separate structure or a transition region between the 5800~\AA\ complex and the $\lambda6225$ diffuse band.
Despite the opposite slopes of the continua in this region (Fig.~\ref{fig:fig1}), on the small wavelength range of the band  the spectrum in the nebula and the re-scaled background follow similar variations (middle left panel of Fig.~\ref{fig:fig7}).

$\lambda6225$,  $\sim100$~\AA\ wide, begins approximatively at 6180~\AA\ and ends at  $\sim6280$~\AA.
It is limited by the OI $\lambda6300$ atmospheric emission line on its red side.
As for $\lambda6050$, the spectrum of the nebula, in this region, reproduces  the pattern of the background (Fig.~\ref{fig:fig7}, middle right plot).
\subsection{$\lambda6380$ and  $\lambda6617$ } \label{seb}
In SCM80, the $\lambda6380$  complex is separated into three emission lines, $\lambda 6361.2$ ($\lambda6361$ in the following),   $\lambda 6378.1$ ($\lambda6378$), $\lambda 6393.7$ ($\lambda6394$).

$\lambda6361$ is on the edge of an atmospheric OI emission line (Fig.~\ref{fig:fig7}, bottom left plot, and Fig.~\ref{fig:fig8}), which renders a comparison with the background difficult.
$\lambda6361$  is furthermore not detected in \cite{winckel02}.

 $\lambda6378$ and $\lambda6394$  are, in ESO observations, broad-band fluctuations of the continuum  (Fig.~\ref{fig:fig8}), centered at 6379 and 6398.8~\AA\  \cite{winckel02}.
$\lambda6378$ slowly diminishes from $s359$/$s360$, in the nebula, towards the edges of the slit but is still present in background spectra (Fig.~\ref{fig:fig8}).
$\lambda6394$ is perceived only over very few pixels (at most from $ s356$ to $s416$, $\sim16''$ on the sky, see Fig.~\ref{fig:fig8}) in the nebula.
It then becomes undetectable.
In the vicinity of the band  the same alternation of maxima and minima may exist in and out of the nebula (Fig.~\ref{fig:fig8}).

The relatively strong $\lambda6617$  band (Fig.~\ref{fig:fig7}, bottom right plot) does not have a counterpart in the background.
The band can be divided into several sub-features, with minima and maxima which may be present in the background spectrum (there is no certainty on this point however).
In ESO observation $11''$ from HD44179 it is observed from, at least, pixel $320$ to pixel $ 445$ (Fig.~\ref{fig:fig9}), thus over $ 125$~pixels on the slit  ($33.5''$ on the sky).
This is slightly larger than the extent of the nebula ($s330$ to $s429$, $27''$) found in Sect.~\ref{neblim}.
\begin{figure}
\resizebox{1.2\columnwidth }{!}{\includegraphics{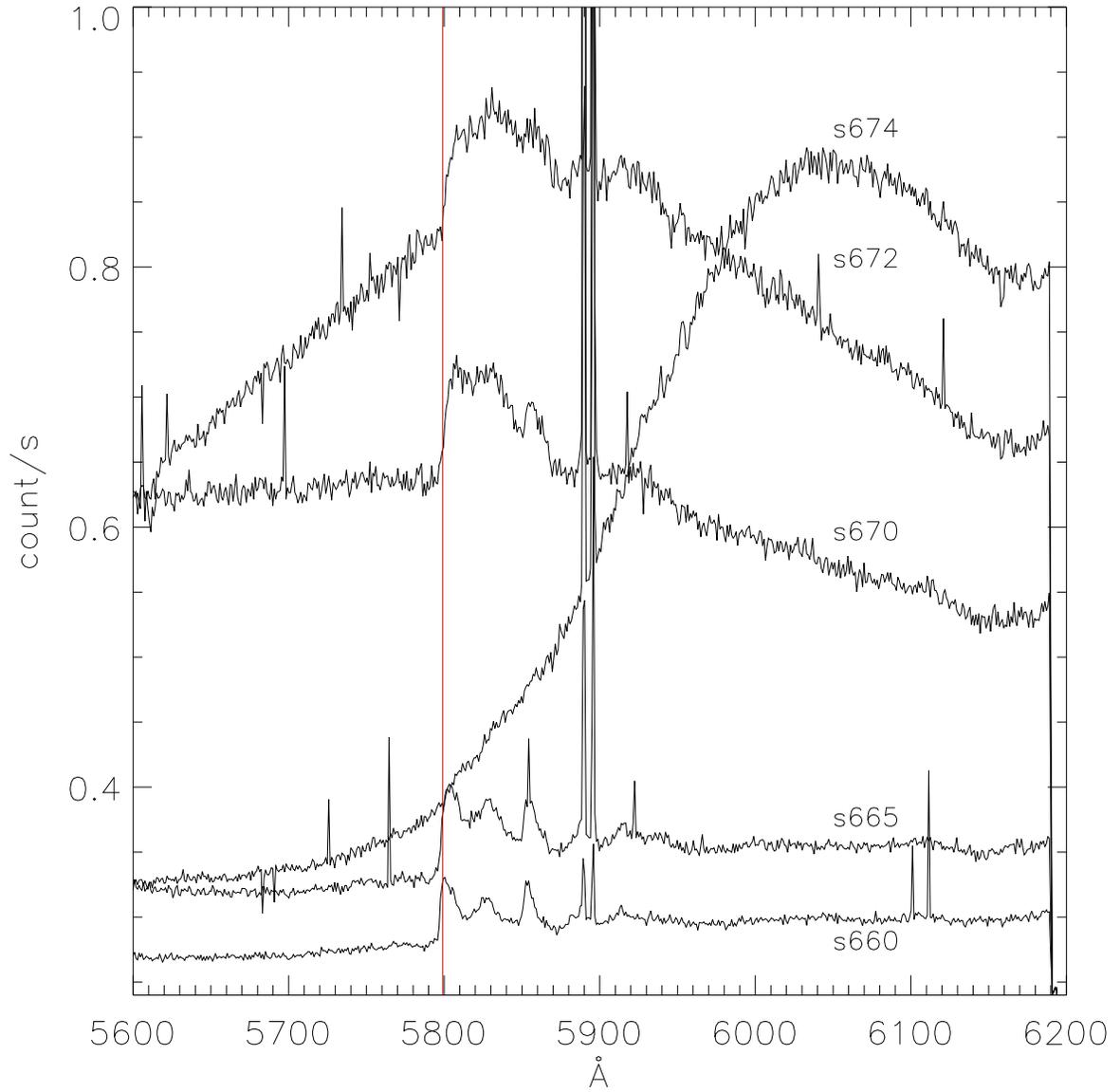}} 
\caption{Variation of the $\lambda5800$  complex with distance from HD44179, from ESO observation along the north-eastern spike of the nebula (P.A.=45$^{\circ}$).
Distance from HD44179 increases with decreasing pixel number.
The continuum of the upper spectra, $s674$, $s672$, $s670$ is dominated by HD44179's diffracted light.
} 
\label{fig:fig10}
\end{figure}
\begin{figure*}
\resizebox{1.3\textwidth }{!}{\includegraphics{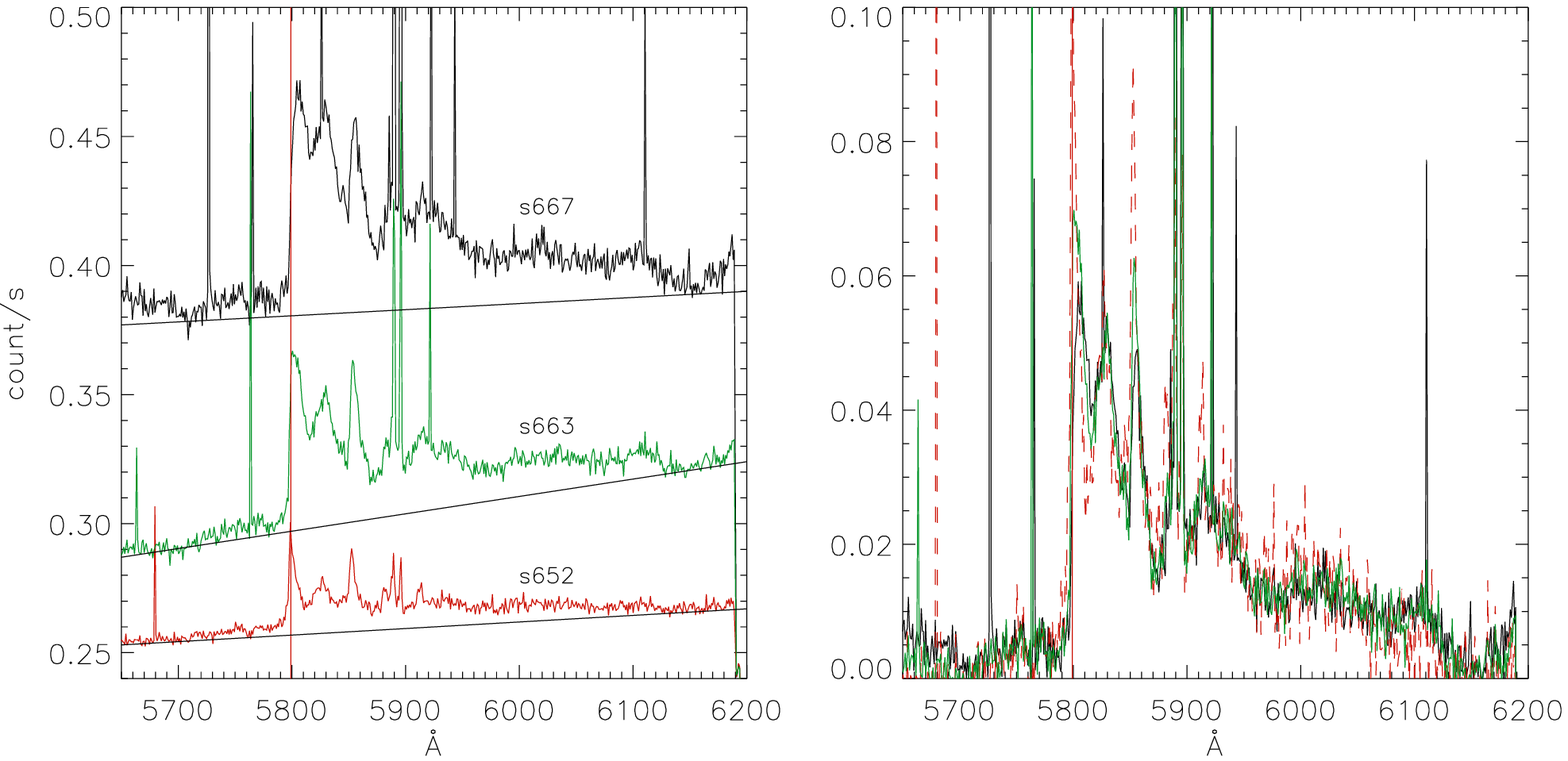}} 
\caption{
Comparison of  spectra from fig.~7 in \cite{winckel02}, before and after re-scaling.
\emph{Left:} $s667$ (top spectrum, $\sim 2.5''$ from the direction of  HD44179), $s663$ (middle spectrum, $1''$  from $s667$ on the slit)  and $s652$ (bottom spectrum, $4''$ from $s667$ ) from ESO observation with P.A.=$45^{\circ}$.
\emph{Right:} the spectra superimpose well after subtraction of a baseline (left plot) and  re-scaling.
} 
\label{fig:fig11}
\end{figure*}
\subsection{The $\lambda 5800$ complex} \label{5800c}
The unidentified SCM80 feature which in the past has attracted most attention is the $\lambda 5800$ complex, divided into six components: $\lambda 5799.2\,\rm\AA$ (F1 in  \cite{schmidt91}), $\lambda 5826.5$ (F2),  $\lambda 5852.9$ (F3), $\lambda 5881.1$ (F4), $\lambda 5912.1$, and $\lambda 5937.0$ (F5).
These central wavelengths are indicative: they vary by a few \AA\ with distance from HD44179, and from paper to paper.

Fig.~\ref{fig:fig10} (ESO observation along the N-E spike of the nebula) shows how the complex's components progressively appear as angular distance  from HD44179 increases.
Starting from the edge of the slit close to HD44179 a vague broad bump is first detected on the top of diffracted/scattered (in the atmosphere) light from the star, in spectrum $s673$.
The relative importance of the bump grows ($s673$ to $s671$) with decreasing pixel number and the lines are progressively individually distinguished ($\sim s665$).
The spectacular decrease of the continuum underlying the $\lambda 5800$ complex observed in this series of spectra is due to the quick attenuation of diffracted starlight.

\begin{figure}
\resizebox{\columnwidth }{!}{\includegraphics{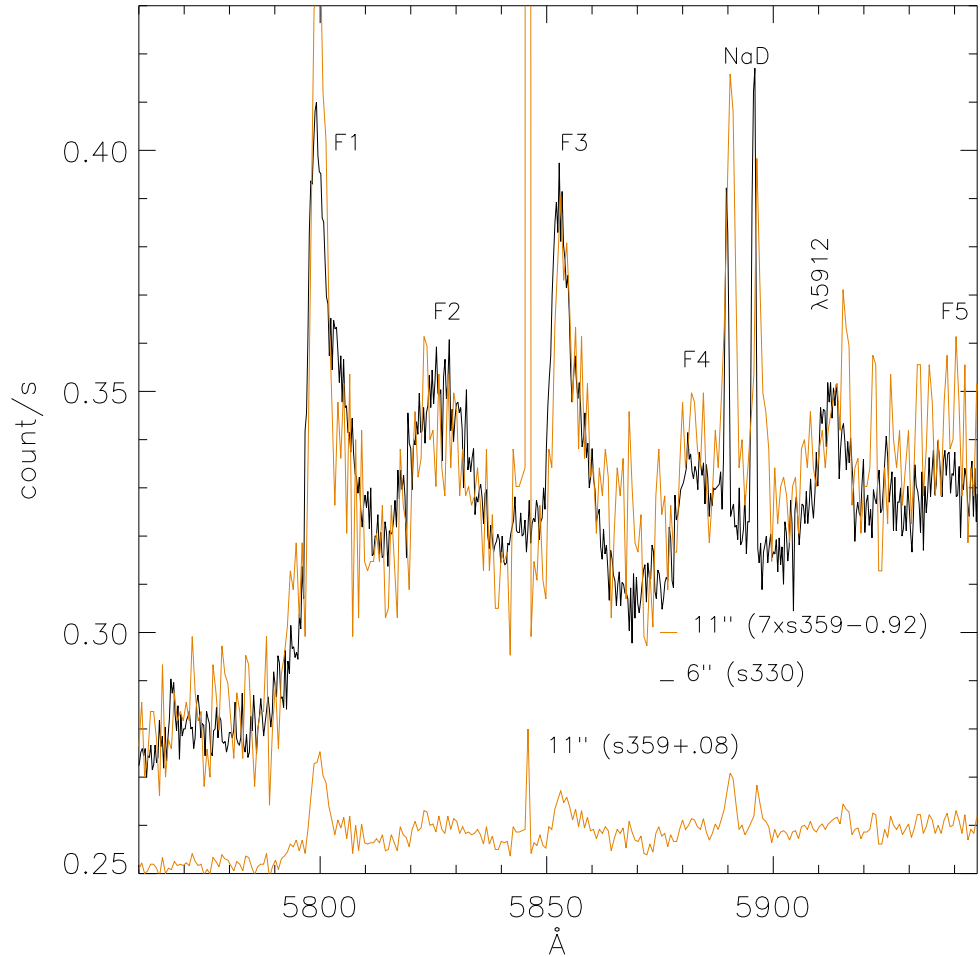}} 
\caption{The $\lambda5800$ SCM80 complex.
Black curve  is the main spectrum ($s330$) of ESO observation $6''$ from HD44179.
Orange bottom curve is $11''$ from HD44179 observation's main spectrum ($s359$), shifted by +0.08 count/s.
After re-scaling of $s359$, the spectra superimpose well.
} 
\label{fig:fig12}
\end{figure}

As it is found in \cite{winckel02,glinski02,schmidt91}, spectra close to the star have the  blue-edges of their  $\lambda 5799.2\,\rm\AA$ and $\lambda 5852.9$ lines red-shifted (Figs.~\ref{fig:fig10} and \ref{fig:fig11}).
However, spectra of the $5800\,\rm\AA$ complex observed at different distances from HD44179 superimpose well, provided that an appropriate baseline be subtracted and after re-scaling (Figs.~\ref{fig:fig11} and \ref{fig:fig12}).
These relationships do not appear in fig.~7 of \cite{winckel02} where the spectra continua are scaled to a common level, with no baseline subtraction.
Since the continuum underlying the complex has no much signification (Sect.~\ref{fdet}), it may be that the bands decrease in proportion of distance from HD44179. 
The observed red-shift of their blue-edge (and the shift in central wavelength)  may thus partly be due to the slope of the underlying continuum.

Figs.~\ref{fig:fig6} (right plot) and \ref{fig:fig7} (top right plot) show that the continua underlying the bands  remain very much alike in the nebula and in the background. 
Two SCM80 bands, $\lambda 5912.1$ and  $\lambda 5937.0$,  are present in the background (Fig.~\ref{fig:fig7}; the two lines also seem to be present in the night sky spectrum of \cite{massey00}). \cite{hanuschik03} finds a telluric emission line at $5915\,\rm\AA$, close to  $\lambda 5912.1$.

The first four bands of the complex rapidly decrease along the slit  from the main pixel of an observation towards the edges of the slit.
In ESO observation $6''$ from HD44179 these lines are present (Fig.~\ref{fig:fig6}) in individual spectra from $\sim s308$ to $\sim s370$ (over $\sim 18''$).
Either by averaging a few spectra or by using a median filter, they are  still  detected from  $\sim s295$ to $\sim s395$ (over $\sim 27''$).
Farther away on the slit the uncertainty on the spectra is too large to reach a conclusion.
In ESO observation $11''$ from HD44179 the lines are present over the same extent ($\sim 27''$), from $s325$ to $s435$, their presence being difficult to ascertain outside this interval.
\section{Discussion} \label{dis}
\subsection{Spatial distribution of the Red Rectangle features} \label{orig}
I have, in Sect.~\ref{neblim}, studied the transition between nebular and background spectra from changes in the overall shape of the spectra along the slit in different observations.
In Sect.~\ref{cfn} a similar work was done separately for each Red Rectangle band.
It appears  that  exact limits of the nebula at a given distance from HD44179 are difficult to determine, because of the resemblance between background and nebular spectra.
It is also manifest  that the extent of the ERE sub-features rarely coincide with these limits.
Some of them are even found in the background spectra.

At the edges of the nebula, nebular and background spectra have similar shapes (Sect.~\ref{neblim}).
$11''$ from HD44179, the only concrete sign  of a transition between the inside and the outside of the nebula is the red edge of the ERE bump.
In ESO observation $6''$ from HD44179, this transition is not perceived because of the importance of  HD44179's diffracted light in the spectrum.
Variations in the 5800~\AA\ region would indicate that the nebula is, at this distance from HD44179, as large as it is $11''$ from the star.

Spectral variations over short wavelength intervals are in general identical for the  background and for the nebula.
In the wavelength regions of the three large, diffuse bands $\lambda5600$, $\lambda6050$, $\lambda6225$,  the  structure of the background and the nebula's spectra are the same.
$\lambda5912$ and $\lambda5937$ in the 5800~\AA\ complex,  $\lambda6378$ in the $\lambda6380$ complex, also seem to be present in background spectra.
These bands may not originate in the nebula.

The extent over which the other SCM80 bands are detected is in better agreement with the expected limits of the nebula.
 In ESO observation $11''$ from HD44179, $\lambda6394$ (Sect.~\ref{seb}) is present only over a few pixels in the nebula.
The first four bands in the $\lambda5800$ complex  (Sect.~\ref{5800c}) and $\lambda6617$ (Sect.~\ref{seb})  are respectively observed over $\sim27''$ and  $\sim 32''$ in ESO observation $11''$ from HD44179.
These widths are close to the one found for the ERE red rise.
I have noted however that the $\lambda5800$ complex is seen over the same extent ($27''$) in ESO observation $6''$ from the star, while, according to \cite{winckel02},  the bi-conical nebula is expected to be smaller at this distance from HD44179. 
\subsection{Consequences for the Red Rectangle spectrum} \label{cons}
Over wavelength intervals of a few tens of \AA\ the spectrum of the nebula and the background are modulated  by small variations which often coincide.
These similarities can be explained  if they belong to the spectrum of HD44179, or if they are due to atmospheric extinction.

In the former case, the features are found in the background  because of the presence of light from HD44179 scattered in the atmosphere.

In the latter, they must be absorption features,  since they are in proportion of the continuum (telluric emission lines would be much larger on the re-scaled background spectrum, see sect.~\ref{cfn}).
This hypothesis might be favored since the night-sky spectrum provided by P.~Massey, although it is of lower resolution than the FAST data, follows, over wavelength intervals of a few tens of \AA, the same variations as the background spectrum used in  Figs.~\ref{fig:fig1} and  \ref{fig:fig7}.
\begin{figure}[h]
\resizebox{\columnwidth }{!}{\includegraphics{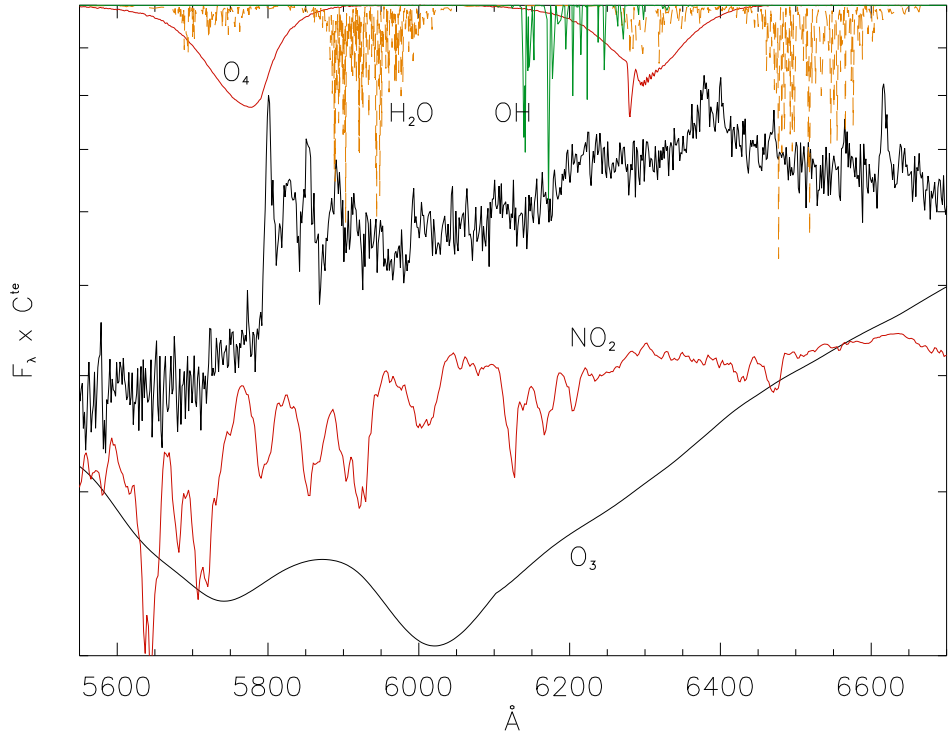}} 
\caption{
The spectrum of the Red Rectangle (FAST observation) of Fig.~\ref{fig:fig1}, low resolution absorption spectra of O$_3$, NO$_2$, O$_4$, H$_2$O and OH.
 y-logarithmic axis.
} 
\label{fig:fig13}
\end{figure}

These similarities are not cancelled  by the data-reduction routines  usually applied to these observations.
Firstly the background to be subtracted to the observed spectrum of the nebula is not the night-sky spectrum: in the blue, in the first few ten arc-seconds from HD44179, the background is dominated by diffracted light or scattered starlight (in the earth's atmosphere) from HD44179 (Sect.~\ref{fdet}).
Secondly, the slight variations due to atmospheric absorption which seem to be present in the spectrum of the nebula will not be removed during the data-reduction process.

Some of the bands found in SCM80 and in the background (Sect.~\ref{cons}) therefore appear to be atmospheric absorption features, or,  features in the spectrum of HD44179.
In fact, the absorption spectrum of a complex molecule can easily be confused with an emission one (Sect.~\ref{bandid}, Figs.~\ref{fig:fig15} and  \ref{fig:fig16}):
both spectra are sequences of alternate 'peaks' and 'valleys'.
\subsection{Atmospheric absorption} \label{atabs}
Fig.~\ref{fig:fig13} plots low resolution  absorption spectra of a few important atmospheric molecules, between 5500 and 6800~\AA, along with the FAST spectrum of the Red Rectangle nebula.
Ozone has a structure no finer than represented on the figure, and is thus probably  not related, except for the ERE, to the SCM80 bands.

Absorption by NO$_2$ peaks around 4400~\AA\ but is present over the whole visible continuum  (Fig.~\ref{fig:fig13});
its complex structure (Figs.~\ref{fig:fig15} and ~\ref{fig:fig16}) is not yet (and may not be) resolved  by observation.

H$_2$O,  O$_4$ (O$_2$-O$_2$), OH....., also have fine structure but operate in limited wavelength ranges.
The spectra of monomer O$_2$ and dimer O$_4$ are intimately mixed in observations of the atmosphere.
O$_2$ has discrete bands at 7633, 6896 and 6289~\AA;
O$_4$ at 6300, 5780 and 4773~\AA.
O$_4$ will preferentially be observed in the lowest atmospheric layers.
\begin{figure}[h]
\resizebox{\columnwidth }{!}{\includegraphics{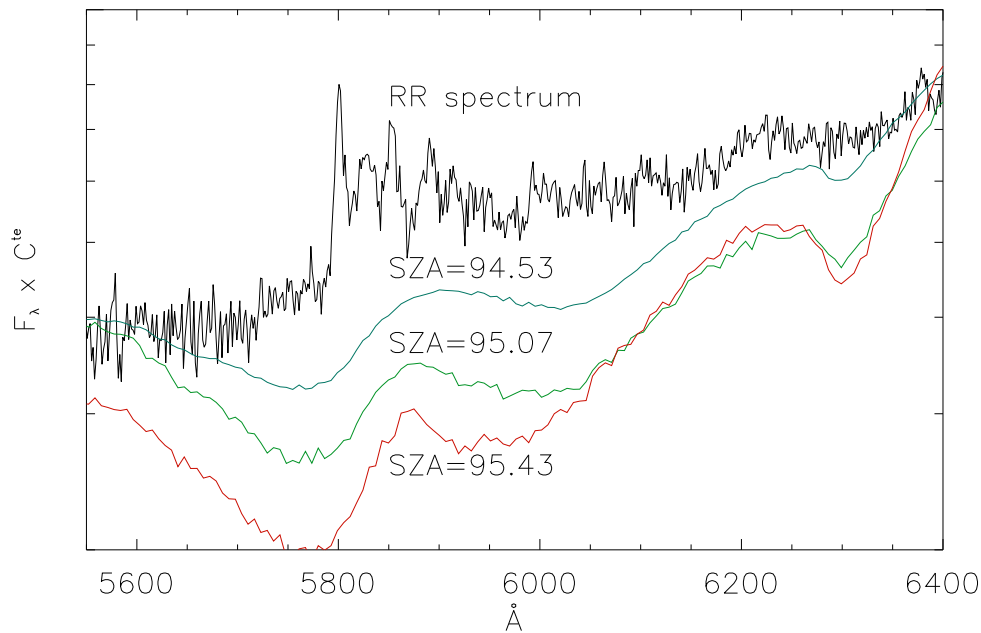}} 
\caption{
Comparison of the Red Rectangle  spectrum of Fig.~\ref{fig:fig1} to SAOZ spectra of a sun occultation.
'SZA'  gives the zenithal angle of the sun (which is below the horizon in these observations).
 y-logarithmic axis.
} 
\label{fig:fig14}
\end{figure}

On Fig.~\ref{fig:fig14}  the FAST Red Rectangle spectrum is compared to spectra of the sun occultation from SAOZ experiment.
The angle (SZA) between the direction of the sun and the zenith is written above each spectrum.

Ozone is responsible for the strong absorption of sunlight observed in the occultation spectra between 5600 and 6700~\AA\ \cite{sol1}.
It may contribute to  the blue decrease of the ERE bump  \cite{rr}.
The bump feature between 5750~\AA\ and 6050~\AA\ on the SZA=94.53 spectrum is also a characteristic of absorption by  ozone (see the absorption spectrum of O$_3$ in Fig.~\ref{fig:fig13}).
As sunlight crosses lower atmospheric layers, O$_4$  provokes the depression at 6300~\AA\ (absorption by O$_2$ and H$_2$O is also present in this region) and shortens the ozone bump on its blue side.
In addition,  the originally smooth aspect of the spectrum at SZA=94.53 gradually breaks because of absorption by other atmospheric molecules (NO$_2$ for instance).
\subsection{Comparison of Red Rectangle and atmospheric absorption spectra} \label{bandid}
\begin{figure}
\resizebox{\columnwidth }{!}{\includegraphics{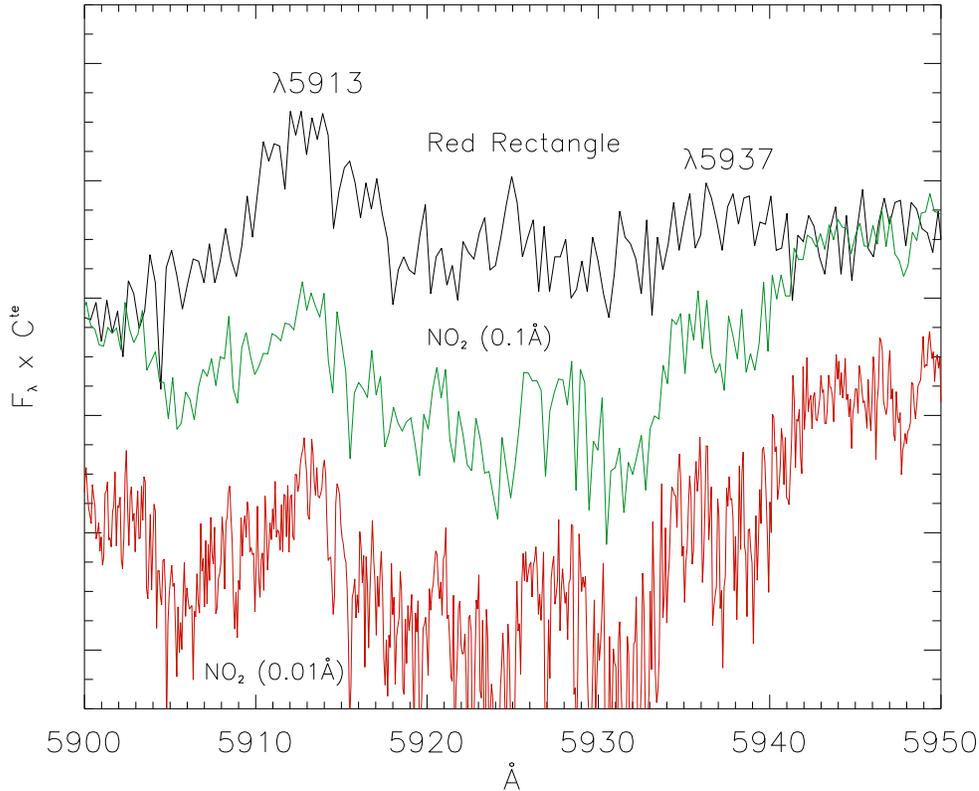}} 
\caption{
Comparison of ESO Red Rectangle  spectrum, $11''$ north from HD44179, to absorption spectra of NO$_2$ (two observations with different resolutions: 0.1 and 0.01~\AA) in the [5900, 5950]\AA\ region.
Note the resemblance between the Red Rectangle spectrum and the absorption spectra of NO$_2$ at \AA\ scales.
} 
\label{fig:fig15}
\end{figure}
\begin{figure}[h]
\resizebox{\columnwidth }{!}{\includegraphics{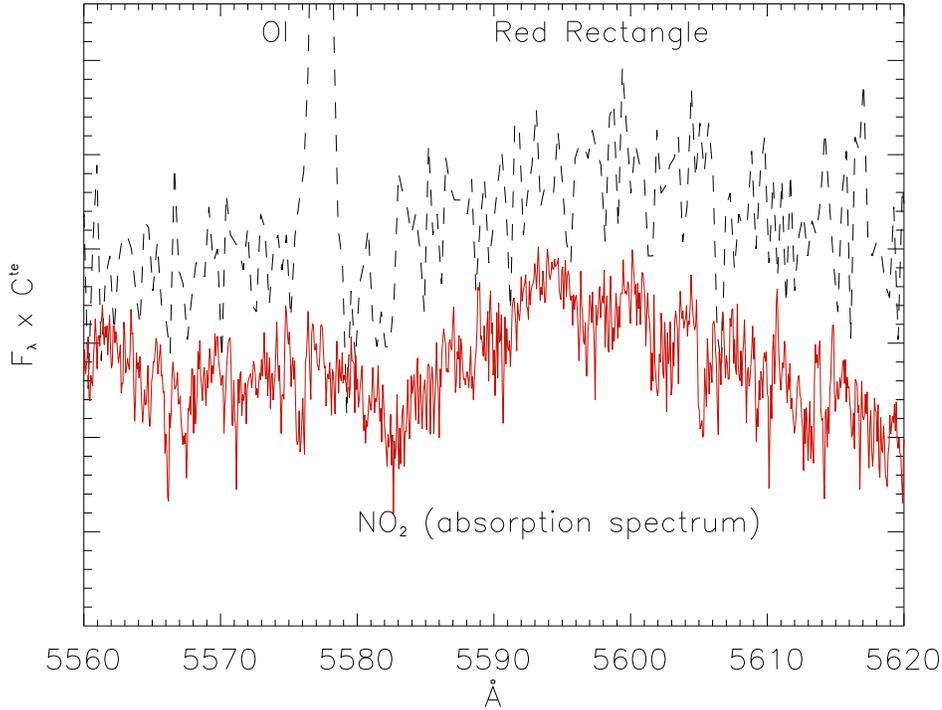}} 
\caption{
Comparison of an ESO Red Rectangle spectrum, $6''$ from HD44179, to the absorption spectrum of NO$_2$  in the 5600~\AA\ region.
The $\lambda5600$ SCM80 diffuse band extends approximately from 5580~\AA\ to 5612~\AA.
In this wavelength region the two spectra follow similar variations.
} 
\label{fig:fig16}
\end{figure}
Figs.~\ref{fig:fig13} and \ref{fig:fig14} show similarities between the low resolution SAOZ spectra, the spectra of atmospheric molecules, and the spectrum of the nebula, which would worth further investigations.

The bump-like continuum which underlies the $\lambda5800$ Red Rectangle complex and was shown to be common to background and nebular spectra (right plot of Fig.~\ref{fig:fig6}), for instance, coincides with the ozone singularity between 5600 and 6700~\AA.

The depression on the red side of Red Rectangle band  $\lambda6225$ corresponds to  O$_4$ and O$_2$ absorption (Fig.~\ref{fig:fig14}).

At higher resolution, I have found a possible correlation (Fig.~\ref{fig:fig15}) between  the  $\lambda5912.1$ and $\lambda5937$ SCM80's bands (Sect.~\ref{5800c}) and A.~Jenouvrier's laboratory spectrum of NO$_2$. 
The $\lambda5600$ diffuse band may also be due to NO$_2$ (Fig.~\ref{fig:fig16}).
\subsection{Night sky} \label{ns}
The characteristic pattern of Kitt Peak and Mount Hopkins night sky spectrum found in \cite{ massey00,massey90} also exists at La Silla and can be thought to hold good in other parts of the world.
It might be representative of a clear atmosphere, and the consequence of atmospheric extinction on general background light.

 \cite{massey00} attributes the $[5500, 6500]$~\AA\ region of the spectrum to city light pollution caused by high-pressure sodium lamps (HPS; for an HPS spectrum see  fig.~3 in \cite{osterbrock76}, fig.~1c in \cite{massey90}).
If so, variations of the bump underlying this region with azimuth and elevation should be observed (it should in particular disappear in observations towards the zenith), while \cite{massey90} finds it is the same in all directions.
It further exists under the NaD lines (and in the same proportion, Fig.~\ref{fig:fig4}) in the background spectra of ESO observations as well as in the SAOZ occultation  of the Sun.

It might, in fact, be caused by the singularity in ozone absorption (Fig.~\ref{fig:fig13}), mentioned in the preceding section.
This  explains the similarity (Figs.~\ref{fig:fig1}, \ref{fig:fig2} and right plot of Fig.~\ref{fig:fig6}) between night sky, background and nebular continua in the $[5600,\,6200]$~\AA\ region.
\section{Conclusion} \label{conc}
This study has continued a previous work which highlighted the importance of atmospheric extinction in the observed  bump-shape spectrum of the Red Rectangle nebula. 
A more detailed comparison of  background and Red Rectangle nebula spectra,  shows that part of the structure found on the blue side of the ERE bump is still related to the earth's atmosphere and may result from absorption, rather than from emission.
For some Red Rectangle bands I was able to propose known atmospheric molecules as the possible carriers of this absorption.
Features common to the background and to the nebula may also correspond to fine structure in HD44179's spectrum.

The particular shape of the night sky continuum was found to be the same in different observatories and may be a constant.
This shape decreases with wave-number,  maybe  because of the action of atmospheric extinction on background light, and shows a bump-like feature between 5800 and 6000~\AA\ due to a singularity in the ozone absorption cross-section.
High resolution studies of the night sky continuum should reveal the absorption by other atmospheric molecules (as NO$_2$).

The night sky spectrum is deeply modified in the vicinity of a star, which makes determination and subtraction of the background  in the observation of nebulae  particularly difficult tasks.
Standard data-reduction routines also seem to fail to remove slight atmospheric absorption effects which are observed in the spectrum of faint extended objects, such as the Red Rectangle nebula.

The possibility that instrumentation contributes to some of the effects met in the course of this study (for instance, the 'diffracted' starlight observed over several pixels of the 2-D arrays could in part be due to scattering within the spectrometer) was not investigated here and should also be considered in future studies. 
\section*{Acknowledgments}
Atmospheric data needed for this study have been obtained thanks to  A.~Jenouvrier, F.~Goutail, R.~McPheat, L.~Rothman.

I am indebted to  D.~Crisp, W.~Hocking, A.~Jenouvrier, J.~Orphal, for exchanges on the effects of atmospheric absorption and turbulence.

This research has used observations made with ESO NTT Telescope at  La Silla-Paranal  Observatory under programme ID 60.c-0473.

{}

\end{document}